\documentstyle[11pt,aaspp4]{article}

\begin{document}

\title{The Spectroscopic Orbit of the Evolved Binary HD 197770}

\author{Karl D.\ Gordon\altaffilmark{1}, Geoffrey C.\ Clayton\altaffilmark{1},
   Tracy L.\ Smith\altaffilmark{2}, Jason P.\ Aufdenberg\altaffilmark{3}, 
   John S.\ Drilling\altaffilmark{1}, Margaret M.\ Hanson\altaffilmark{4,6},
   Christopher M.\ Anderson\altaffilmark{5}, 
   \& Christopher L.\ Mulliss\altaffilmark{2}}
\altaffiltext{1}{Department of Physics \& Astronomy, 
   Louisiana State University, Baton Rouge, LA 70803}
\altaffiltext{2}{Ritter Astrophysical Research Center, 
   The University of Toledo, Toledo, OH 43606}
\altaffiltext{3}{Department of Physics \& Astronomy, 
   Arizona State University, Tempe, AZ 85271}
\altaffiltext{4}{Steward Observatory, University of Arizona,
   Tucson, AZ 85721}
\altaffiltext{5}{Space Astronomy Laboratory, University of Wisconsin,
   Madison, WI 53706}
\altaffiltext{6}{Hubble Fellow}

\lefthead{Gordon et al.}
\righthead{HD 197770}

\begin{abstract}
  We have used spectra taken between 1992 and 1997 to derive the
spectroscopic orbit of the eclipsing double-lined spectroscopic binary
HD~197770.  This binary has a period of $99.69 \pm 0.02$~days and K
amplitudes of $31.2 \pm 0.8$ and $47.1 \pm 0.4$~km~s$^{-1}$ for
components A \& B, respectively.  The $m\sin^3i$ values for A \& B are
2.9 and 1.9, respectively, and are close to the actual masses due to
the eclipsing nature of this binary.  Both components of HD~197770
have spectral types near B2~III.  This means both components are
undermassive by about a factor of five and, thus, evolved stars.
Additional evidence of the evolved nature of HD~197770 is found in 25,
60, and 100~\micron\ IRAS images of HD~197770.  These images show 2
apparent shells centered on HD~197770; a bright 60~\micron\ shell with
a $14\arcmin$ diameter and a larger ($1\fdg 2$ diameter) bubble-like
feature.  At least one of the components of HD~197770 is likely to be
a post-AGB star.
\end{abstract}

\keywords{stars: individual (HD 197770) -- binaries: spectroscopic}

\section{Introduction}

  Interest in the star HD~197770 (HR~7940, $\alpha(2000) = 20^{\rm
h}43^{\rm m}13\fs 52$, $\delta(2000) = +57\arcdeg 6\arcmin 50\farcs
9$) increased greatly with the discovery that its line-of-sight has a
polarization feature coincident with the 2175~\AA\ extinction bump
(\cite{cla92}; \cite{and96}).  Out of the 30 sightlines with UV
spectropolarimetry, such a polarization bump has only been seen along
one other sightline (\cite{wol97}).  HD~197770 has long been known to
have a variable radial velocity (\cite{ada24}; \cite{pla30}).
Observations originally intended to study the sightline towards
HD~197770 have shown it to be a double-lined spectroscopic binary
(\cite{han94}; \cite{cla96}).  Recent photometric observations have
shown that HD~197770 is an eclipsing binary (\cite{jer93};
\cite{cla96}).  We obtained spectra of HD~197770 between 1992 and 1997 in
order to determine the spectroscopic orbit of HD~197770.  The
spectroscopic orbit, coupled with the eclipsing nature of the binary,
allowed us to determine the masses of the binary components.

\section{Observations}

  In 1992 June and October, observations of HD~197770 were acquired at
the Kitt Peak National Observatory (KPNO) 0.9m coud\'e feed telescope
using the coud\'e spectrograph.  The June observation was obtained
using an \'echelle grating and a cross disperser grism to give
disjoint orders between 3870 and 4085~\AA\ with a resolution of
120,000.  It was reduced using IRAF.  The October observations also
used an \'echelle grating and a cross disperser grism giving orders
which covered 5580--7160~\AA\ (HJD 2448901.6) or 3990--4490~\AA\ (HJD
2448902.6 \& 2448903.6) at a resolution of 80,000.  Between 1996
September and 1997 October, 17 observations of HD~197770 were taken at
the 1m Ritter Observatory telescope using a fiber-fed \'echelle
spectrograph (\cite{gor97}).  These observations cover nine disjoint
70~\AA-orders between 5200 and 6600~\AA\ at a resolution of 25,000.
The KPNO and Ritter observations were reduced using a package written
to reduce Ritter observations.  Since the KPNO observations were not
acquired with a fiber-fed \'echelle, the orders were extracted using a
direct sum method instead of the profile-weighted method used for the
Ritter observations.  Details of the reduction package can be found in
Gordon \& Mulliss (1997).  The UT date and time, exposure length, and
heliocentric Julian Date (HJD) for the observations are listed in
columns 1-4 of Table~\ref{table_obs}.

\begin{deluxetable}{ccrccrrrr}
\tablewidth{0pt}
\tablecaption{Observations and Radial Velocites\label{table_obs}}
\tablehead{ & & \colhead{exp.} & & & \multicolumn{2}{c}{A} &
           \multicolumn{2}{c}{B} \\
           \colhead{UT Date} & \colhead{UT time} & 
           \colhead{time} & \colhead{HJD - 2400000} & \colhead{phase} &
           \colhead{O} & \colhead{O-C} & \colhead{O} & \colhead{O-C} \\
           \colhead{[yy/mm/dd]} & \colhead{[hrs:min]} &
           \colhead{[sec]} & \colhead{[days]} & & 
           \multicolumn{2}{c}{[km s$^{-1}$]} & 
           \multicolumn{2}{c}{[km s$^{-1}$]}}
\startdata
92/06/07 & 7:10 &  4018 & 48780.798 & 0.957 &  -0.4 &  -7.3 & -52.2 &  -3.4 \nl      
92/10/06 & 2:30 &  5400 & 48901.607 & 0.168 & -41.2 &  -1.5 &  21.9 &   0.4 \nl      
92/10/07 & 3:05 & 14400 & 48902.630 & 0.179 & -39.3 &   1.5 &  28.8 &   5.6 \nl      
92/10/09 & 2:42 &  5400 & 48904.614 & 0.199 & -43.1 &  -0.5 &  24.7 &  -1.2 \nl      
96/09/19 & 4:03 &  3600 & 50345.671 & 0.654 & \nodata & \nodata & -31.4 &   4.4 \nl
96/09/20 & 3:10 &  3600 & 50346.634 & 0.663 & \nodata & \nodata & -39.6 &  -1.6 \nl
96/09/25 & 4:33 &  3600 & 50351.692 & 0.714 & \nodata & \nodata & -52.2 &  -3.5 \nl
96/10/01 & 1:38 &  3600 & 50357.570 & 0.773 & \nodata & \nodata & -57.4 &   1.5 \nl  
96/10/04 & 3:12 &  3600 & 50360.636 & 0.804 & \nodata & \nodata & -51.2 &  11.4 \nl
96/10/29 & 1:40 &  3600 & 50385.571 & 0.054 & \nodata & \nodata &  -9.8 &   3.0 \nl
96/11/04 & 0:32 &  3600 & 50391.524 & 0.114 & \nodata & \nodata &   6.8 &  -1.4 \nl
96/11/15 & 1:53 &  3600 & 50402.580 & 0.225 & \nodata & \nodata &  27.4 &  -0.8 \nl
96/12/21 & 0:32 &  3600 & 50438.522 & 0.585 & \nodata & \nodata & -24.2 &  -4.0 \nl
97/08/01 & 6:02 &  3600 & 50661.753 & 0.824 & \nodata & \nodata & -64.6 &  -0.5 \nl
97/08/03 & 6:07 &  3600 & 50663.756 & 0.844 & \nodata & \nodata & -63.6 &   1.1 \nl  
97/08/05 & 7:54 &  3600 & 50665.831 & 0.865 & \nodata & \nodata & -62.4 &   2.0 \nl
97/08/10 & 5:36 &  2400 & 50670.735 & 0.914 & \nodata & \nodata & -59.2 &  -0.2 \nl  
97/09/04 & 3:26 &  3600 & 50695.645 & 0.164 & \nodata & \nodata &  17.0 &  -3.7 \nl
97/09/19 & 3:22 &  3600 & 50710.642 & 0.315 & \nodata & \nodata &  24.4 &  -2.8 \nl
97/09/26 & 3:10 &  3600 & 50717.634 & 0.385 & \nodata & \nodata &  21.2 &   1.9 \nl
97/10/06 & 2:59 &  3600 & 50727.626 & 0.485 & \nodata & \nodata &   4.8 &   3.2 \nl
\enddata
\end{deluxetable}

\section{Analysis \label{sec_analysis}}

\begin{figure}[tbp]
\begin{center}
\plotone{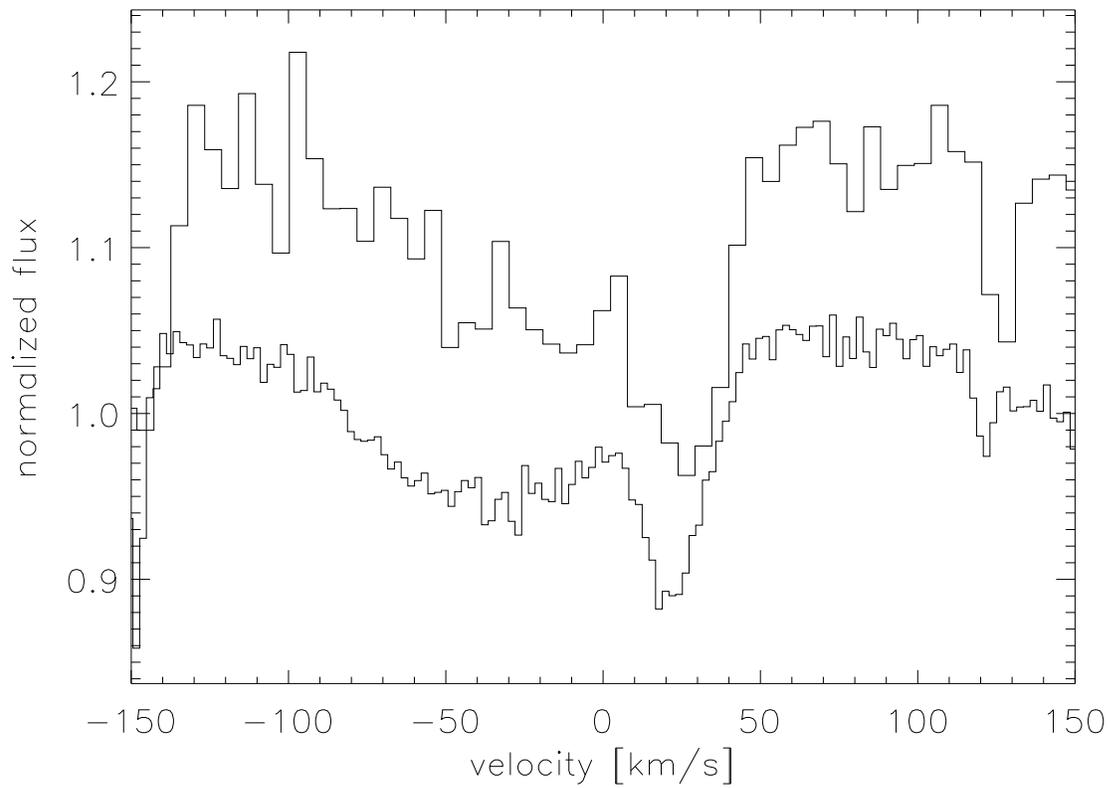}
\caption{The \ion{C}{2} $\lambda$6578.03~\AA\ in the HJD 2448901.6 KPNO
spectrum (bottom) and HJD 2450710.6 Ritter spectrum (top).  The
resolutions of the spectra are 80,000 (bottom) and 25,000
(top). \label{fig_ex_line}}
\end{center}
\end{figure}

  From inspection of the spectra, we found that the lines of the two
components of HD~197770 were never fully separated.
Figure~\ref{fig_ex_line} displays the \ion{C}{2} $\lambda$6578~\AA\
line for typical KPNO and Ritter spectra.  The separation of the
components in the two spectra are similar and the double-lined nature
of this binary is clearly seen, especially in the KPNO spectrum.  The
lines are clearly resolved in the KPNO spectrum, but not in the Ritter
spectrum.  We identify the broad line as component A of this binary
and the narrow line as component B.  Since the lines were never fully
separated, blending effects are always present.  Therefore, we used
TODCOR, the 2D cross correlation algorithm presented in Zucker \&
Mazeh (1994), to measure the radial velocities.  This algorithm takes
model spectra of the two components as input and finds the optimal
radial velocities of the two components and their luminosity ratio.

\begin{figure}[tbp]
\begin{center}
\plotone{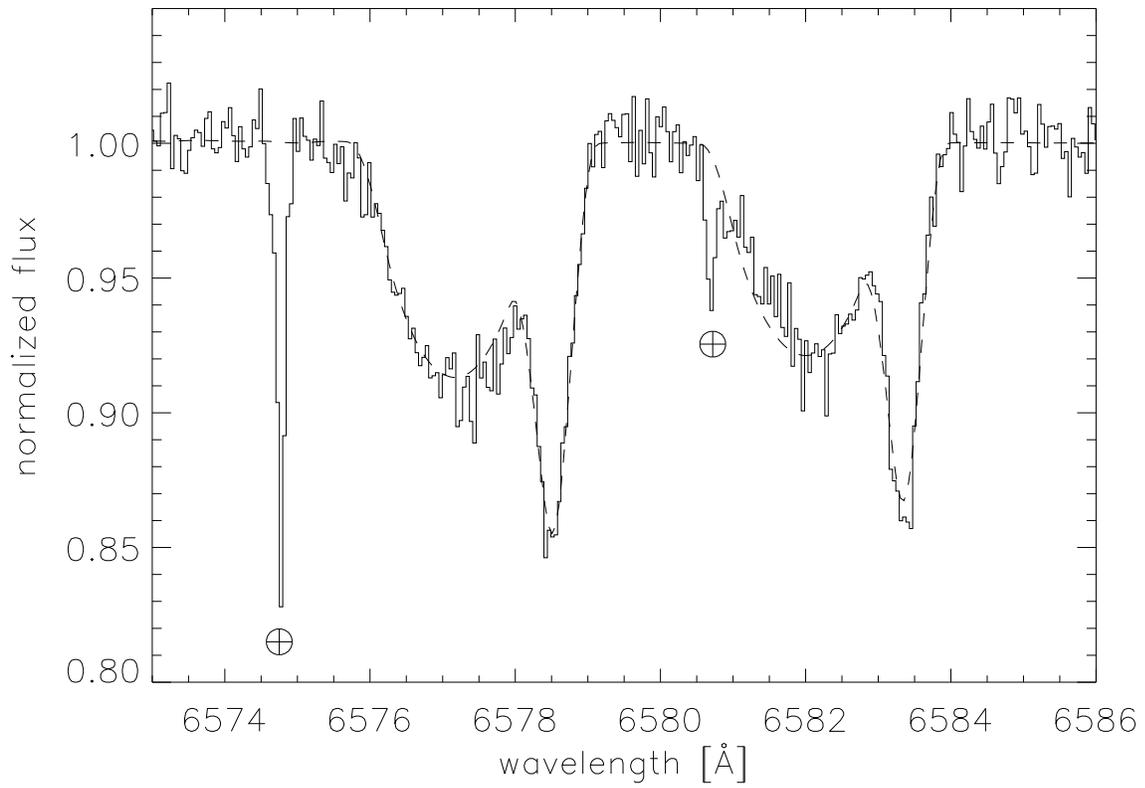}
\caption{The \ion{C}{2} $\lambda\lambda$6578,6582 doublet in the HJD
2448901.6 spectrum is plotted along with the best fit composite model
spectrum. \label{fig_vsini}}
\end{center}
\end{figure}

  As results of TODCOR are sensitive to the assumed model spectra, we
constrained the model atmosphere parameters as follows.  The effective
temperature ($T_{\rm eff}$) for HD 197770, derived using a
reddening-free photometric index, is $21,000 \pm 3,000$ K
(\cite{gul89}).  We checked this $T_{\rm eff}$ by comparing the
ultraviolet through V band dereddened spectral energy distribution
(SED) of HD~197770 with PHOENIX LTE model SEDs (\cite{auf98}).  The
SED of HD~197770 was taken from International Ultraviolet Explorer ({\it
IUE}) and UBV (\cite{har94}) observations and dereddened assuming a
$R_V = 3.1$ and $E(B-V) = 0.58$ (\cite{car89}).  The unreddened SED
was well fit by a model with a $T_{\rm eff} = 21,000$~K and $\log(g) =
4.0$.  In addition, since the eclipse depths are similar, the $T_{\rm
eff}$ of both stars must be close to 21,000~K (\cite{cla96}).  We used
a PHOENIX LTE model SED with a $T_{\rm eff} = 21,000$~K and a $\log(g)
= 4.0$ for both stars for the TODCOR algorithm.  The $v \sin i$ values
were determined by iteratively running TODCOR and adjusting the $v
\sin i$ values until the composite model spectrum matched, by
inspection, the \ion{C}{2} $\lambda\lambda$6578,6582 doublet in the
HJD 2448901.6 spectrum ($R \sim 80,000$).  The strength of the model
spectrum \ion{C}{2} lines were reduced by 40\% to match the observed
strengths of the lines.  This can be traced to the fact that
the model spectrum had solar abundances and most B-type stars are
underabundant compared to the Sun (\cite{sno96} and references
therein).  The HJD 2448901.6 spectrum and the best fitting composite
model spectrum is shown in Figure~\ref{fig_vsini}.  The best fit
$v\sin i$ values were 55 and 15~km~s$^{-1}$ for components A and B,
respectively.  The TODCOR determined luminosity ratio ($L_A/L_B$) was
$\sim$2.

\begin{figure}[tbp]
\begin{center}
\plotone{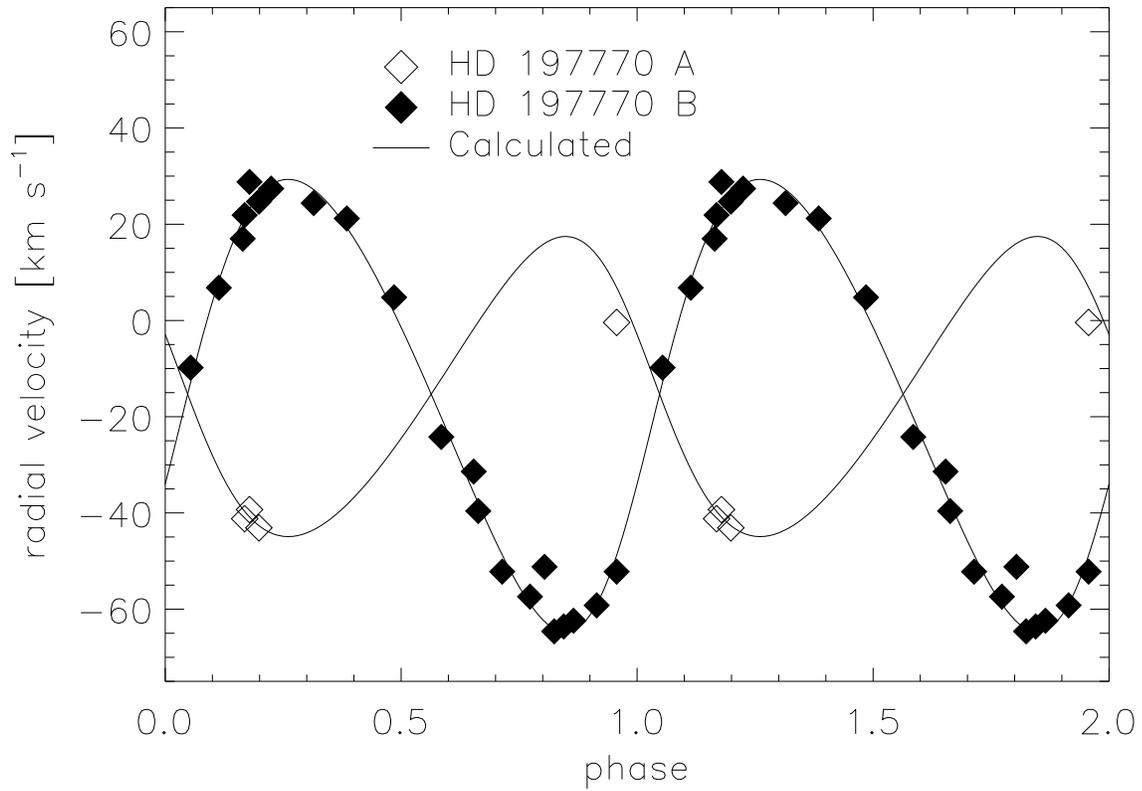}
\caption{The radial velocities for HD~197770 A \& B are plotted phased
to the period given in Table~2.  The radial velocities calculated from
the fits to the orbital motion are plotted as solid
lines. \label{fig_orbit}}
\end{center}
\end{figure}

  Using TODCOR, we were able to measure the radial velocity of the
narrow, but not the broad, component in all of the spectra.  We were
able to accurately measure the broad component in only the 1992
October KPNO spectra.  The 1992 June KPNO spectrum gave us radial
velocities for both components, but neither was measured with great
accuracy due to the low signal-to-noise ratio (S/N).  The resolution
and S/N of the Ritter spectra were not high enough to allow us to
measure the velocity of the broad component (see
Fig.~\ref{fig_ex_line}).  As a result, the orbital motion of the
narrow component was well sampled, but the orbital motion of the broad
component was only measured 4 times, 3 times accurately.  We fit the
narrow component radial velocities using all but the HJD 2450360.6
measurement.  This one radial velocity is quite deviant ($4\sigma$
from final fit, $\sigma = 2.8$~km~s$^{-1}$) and a much better fit is
achieved by rejecting this point.  We fit the 3 good measurements of
the broad component to determine $K_A$.  For this fit, we assumed all
the other orbital parameters were those determined from the narrow
component fit, except $w_A = w_B - 180\arcdeg$.  The resulting orbital
parameters are listed in Table~\ref{table_orbit}.  The values of $m
\sin^3 i$ and $a \sin i$ were then computed from the orbital
parameters. Figure~\ref{fig_orbit} plots the measured radial
velocities along with the radial velocities calculated from the
orbital motion fits.  Columns 5-9 of Table~\ref{table_obs} give the
phase, measured radial velocites (O), and observed minus calculated
(O-C) radial velocities.

\begin{deluxetable}{ccc}
\tablewidth{0pt}
\tablecaption{Orbital Parameters \label{table_orbit}}
\tablehead{ & \colhead{A} & \colhead{B}}
\startdata
$V_o$ [km s$^{-1}$] & \multicolumn{2}{c}{$-15.31 \pm 0.32$} \nl
$K$ [km s$^{-1}$] & $31.21 \pm 0.81$ & $47.05 \pm 0.36$ \nl
$e$ & \multicolumn{2}{c}{$0.147 \pm 0.008$} \nl
$w$ [\arcdeg] & $69.76 \pm 3.34$ & $249.76 \pm 3.34$ \nl
$P$ [days] & \multicolumn{2}{c}{$99.6918 \pm 0.0223$} \nl
$T_o$ [days] & \multicolumn{2}{c}{$41906.392 \pm 1.975$} \nl
$m \sin^3 i$ [M$_{\sun}$] & $2.89 \pm 0.08$ & $1.92 \pm 0.09$ \nl
$a \sin i$ [R$_{\sun}$] & $60.8 \pm 1.6$ & $91.7 \pm 0.7$ \nl
\enddata
\end{deluxetable}

\section{Discussion}

  Since HD~197770 exhibits shallow eclipses ($\delta V \sim 0.05$),
its inclination must be near $90\arcdeg$ (\cite{cla96}).  Thus, the
values quoted in Table~\ref{table_orbit} for $m\sin^3 i$ are close to
the actual masses of the components.  Comparing an unpublished Pine
Bluff Observatory (PBO) spectrum of HD~197770 to the spectra presented
in Walborn \& Fitzpatrick (1990), we find the spectral class to be
B1~V-III or B2~III for both stars combined.  Considering both the PBO
spectrum and the $T_{\rm eff}$ value determined in
\S\ref{sec_analysis}, the most likely spectral type is B2~III.  All 
lines in the spectra of HD~197770 are double, leading to the
conclusion that both stars have similar spectral types.  Assuming
the radii of a B2~III star (12~R$_{\sun}$; \cite{dri97}) for both
components, the inclination of this eclipsing binary is $\geq
73\arcdeg$. The masses of components A \& B are then $\leq 3.3$ and
2.2~M$_{\sun}$, respectively.  A normal B2~III star has a mass around
15 M$_{\sun}$ (\cite{dri97}).  Thus, both components of this binary
are undermassive for their given spectral types.  This marks both
components as evolved stars.

\begin{figure}[tbp]
\begin{center}
\plotone{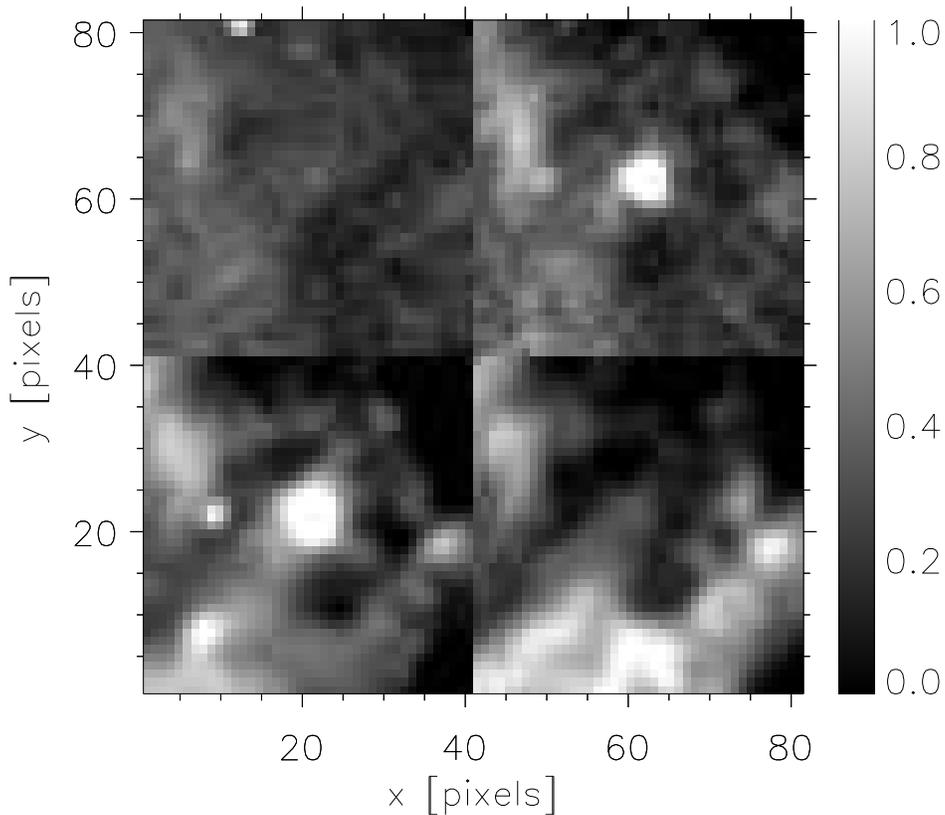}
\caption{The IRAS $2\arcdeg \times 2\arcdeg$ images centered on
HD~197770 are displayed.  The image intensity ranges are
1.2-2.5~MJy~sr$^{-1}$ for the 12~\micron\ image (upper left),
4.2-4.95~MJy~sr$^{-1}$ for the 25~\micron\ image (upper right),
4.0-6.0~MJy~sr$^{-1}$ for the 60~\micron\ image (lower left), and
25-35~MJy~sr$^{-1}$ for the 100~\micron\ image (lower right).
\label{fig_all_four}}
\end{center}
\end{figure}

  HD~197770 lies in the Cygnus region on the edge of Cyg OB7 and Cep
OB2.  It seems to be very near and possibly on the edge of a molecular
cloud/star formation region including Lynds~1036 and 1049.  HD~197770
is associated with two IRAS point sources, IRAS~20420+5655 and
20418+5700, but it is clearly non-stellar on the IRAS map at
60~\micron.  Figure~\ref{fig_all_four} displays the $2\arcdeg \times
2\arcdeg$ region centered on HD~197770 for the IRAS 12, 25, 60,
100~\micron\ bands.  The IR SED peaks at 60~\micron\ implying warm
dust in the immediate vicinity of the binary (\cite{gau93}).  In
addition to the 60~\micron\ shell with a radius of 14$\arcmin$, the
IRAS map shows the signature of an apparent bubble of cleared dust
with a radius of approximately $0\fdg 6$.  This bubble is easiest to
see in the 60 and 100~\micron\ images, but is also present in the
25~\micron\ image.  The dark regions in the upper and lower right hand
corners of the images mark the edge of a molecular cloud which has
been mapped in CO (\cite{dob94}).  The molecular cloud wraps around
three sides of HD~197770 roughly surrounding the evacuated area in
the IRAS map.  In particular, to the west of HD~197770 is a buried
young stellar object (Lynds~1036) and to the east of HD~197770 is a
pulsar (\cite{dew85}). So sequential star formation seems possible
although the pulsar may be a background object. In any event, the
HD~197770 binary appears to have formed on the edge of this cloud and
cleared out an area around it.  The association of HD~197770 with this
cloud allows us to use the distance to the cloud (440~pc;
\cite{dob94}) as an estimate of the distance to HD~197770.

  The existence of IR flux peaking at 60~\micron\ at the position of
HD~197770 and bubble of cleared dust surrounding HD~197770 imply that
this binary has undergone at least two episodes of mass loss.  A
measurement of the luminosity of this binary would greatly help in
determining its evolutionary stage, but the distance to the system is
not known.  HIPPARCOS gives a parallax of $0.52 \pm 0.50$~mas which
results in a $3\sigma$ lower limit on the distance of 495~pc.  This is
consistent with the ``guilt by association'' distance given in the
last paragraph.  By assuming the distance to HD~197770 is 440~pc,
$L_B/L_A = 2$, and $T_{\rm eff} = 21,000$~K, we can estimate the
luminosity and radii of the components of this binary.  The resulting
luminosities and radii are $9.3 \times 10^3$~L$_{\sun}$ ($M_V = -3.2$)
and 7.4~R$_{\sun}$ for HD~197770~A and $4.7 \times 10^3$~L$_{\sun}$
($M_V = -2.5$) and 5.2~R$_{\sun}$ for HD~197770~B.  These radii result
in an inclination for the binary of $\geq 81\arcdeg$ and, thus, masses
of $\sim$3.0 and 2.0~M$_{\sun}$ for HD~197770~A \& B, respectively.

  We note the similarity between this system and the eclipsing binary
$\upsilon$~Sgr ($m_p\sin^3 i = 2.5$~M$_{\sun}$, $m_s\sin^3 i =
4.0$~M$_{\sun}$, \& $a\sin i = 210$~R$_{\sun}$; \cite{dud90}) for
which the primary is a hydrogen-deficient A-type supergiant and the
secondary has a spectral type of B2~Ib (\cite{sch83}).  According to
Plavec (1973) and Schoenberner \& Drilling (1983), the system is the
result of type BB binary evolution, where the primary is now a
post-AGB star.  Uomoto (1986) has discussed the possibility that such
systems are progenitors of Ib-type supernovae.

  Two kinds of additional observations are needed of this binary.
First, high resolution (R~$\geq 50,000$) and high S/N optical spectra
are needed spaced throughout the $\sim$100 day orbit.  These
observations would confirm the K-amplitude of component A which is
currently only based on three measurements.  Second, imaging with an
optical interferometer would give an accurate distance measurement
when coupled to the results of this work.  The Navy Prototype Optical
Interferometer could perform this imaging when it is fully operational
(\cite{arm98}).  Calculation of each star's luminosity would help in
understanding their evolution stage.

\acknowledgments

  The Ritter observations were possible only with the help of the
Ritter technician Bob Burmeister and the crack Ritter observing team.
Team members contributing observations to this effort were Karl
Gordon, David Knauth, Anatoly Miroshnichenko, Nancy Morrison,
Christopher Mulliss, \& Tracy Smith.  Support for observational
research at Ritter Observatory is provided by NSF grant AST-9024802 to
B.\ W.\ Bopp and by The University of Toledo.  This research has made
use of the Simbad database, operated at CDS, Strasbourg, France.  The
IRAS data were obtained using the ISSA postage stamp server at IPAC.
IPAC is funded by NASA as part of the IRAS extended mission under
contract to JPL.  K.\ D.\ Gordon \& G.\ C.\ Clayton acknowledge
support from NASA grant NAS5-3531.  C.\ M.\ Anderson acknowledges
support from NASA grant NAS5-26777.

\end{document}